\documentclass[
reprint, 
prx,
superscriptaddress,
 amsmath,amssymb,
 aps,
 longbibliography
]{revtex4-2}

\let \revappendix \appendix

\usepackage{graphicx}
\usepackage{xcolor}
\usepackage[normalem]{ulem}
\usepackage{braket}
\usepackage{tikz}
\usepackage{changes}
\usepackage{dcolumn}
\usepackage{bm}
\usepackage{siunitx}
\usepackage{microtype}
\usepackage{upgreek}
\usepackage{hyperref}
\usepackage{lipsum}
\hypersetup{
    colorlinks,
    linkcolor={blue!80!black},
    citecolor={blue!80!black},
    urlcolor={blue!80!black}
}

\definecolor{myblue}{RGB}{0, 112, 192}

\DeclareUnicodeCharacter{2212}{-}

\begin{document}

\author{L.~Banszerus}
\affiliation{Center for Quantum Devices, Niels Bohr Institute, University of Copenhagen, 2100 Copenhagen, Denmark}
\author{C. W. Andersson}
\affiliation{Center for Quantum Devices, Niels Bohr Institute, University of Copenhagen, 2100 Copenhagen, Denmark}
\author{W. Marshall}
\affiliation{Center for Quantum Devices, Niels Bohr Institute, University of Copenhagen, 2100 Copenhagen, Denmark}
\affiliation{Department of Physics, University of Washington, Seattle, Washington 98195, USA}%
\author{T. Lindemann}
\affiliation{Department of Physics and Astronomy, Purdue University, West Lafayette, Indiana 47907, USA}%
\affiliation{Birck Nanotechnology Center, Purdue University, West Lafayette, Indiana 47907, USA}
\author{M. J. Manfra}
\affiliation{Department of Physics and Astronomy, Purdue University, West Lafayette, Indiana 47907, USA}%
\affiliation{Birck Nanotechnology Center, Purdue University, West Lafayette, Indiana 47907, USA}
\affiliation{School of Electrical and Computer Engineering, Purdue University, West Lafayette, Indiana 47907, USA}
\affiliation{School of Materials Engineering, Purdue University, West Lafayette, Indiana 47907, USA}
\author{C. M. Marcus}
\affiliation{Center for Quantum Devices, Niels Bohr Institute, University of Copenhagen, 2100 Copenhagen, Denmark}
\affiliation{Department of Physics, University of Washington, Seattle, Washington 98195, USA}%
\affiliation{Materials Science and Engineering, University of Washington, Seattle, Washington 98195, USA}%
\author{S. Vaitiek\.{e}nas}
\affiliation{Center for Quantum Devices, Niels Bohr Institute, University of Copenhagen, 2100 Copenhagen, Denmark}%

\title{The hybrid Josephson rhombus:\\A superconducting element with tailored current-phase relation}

\date{\today}

\begin{abstract} 
Controlling the current-phase relation (CPR) of Josephson elements is essential for tailoring the eigenstates of superconducting qubits, tuning the properties of parametric amplifiers, and designing nonreciprocal superconducting devices.
Here, we introduce the hybrid Josephson rhombus, a highly tunable superconducting circuit containing four semiconductor-superconductor hybrid Josephson junctions embedded in a loop.
Combining magnetic frustration with gate-voltage-controlled tuning of individual Josephson couplings provides deterministic control of the harmonic content of the rhombus CPR.
We show that for balanced Josephson couplings at full frustration, the hybrid rhombus displays a $\pi$-periodic $\cos(2\varphi)$ potential, indicating coherent charge-$4e$ transport.
Tuning away from the balanced configuration, we observe a superconducting diode effect with efficiency exceeding 25\%.
These results showcase the potential of hybrid Josephson rhombi as fundamental building blocks for noise-resilient qubits and quantum devices with custom transport properties.
\end{abstract}

\maketitle

\section{Introduction}
The Josephson effect enables dissipationless current flow between two weakly coupled superconducting leads~\cite{Josephson1962Jul}.
The magnitude and direction of the supercurrent, $I$, depends on the phase difference, $\varphi$, across the Josephson junction~(JJ), characterizing its current-phase relation~(CPR).
Tunneling JJs with insulating barriers (SIS) display sinusoidal CPRs, where $I$ is predominantly carried by sequential tunneling of individual Cooper pairs~\cite{Golubov2004Apr,Willsch2024Feb}.
In contrast, normal-conductor junctions (SNS) with high transparency exhibit a nonsinusoidal CPR containing higher harmonics, ${I(\varphi)=\sum\nolimits_{n=1}^\infty A_n \sin(n \varphi)}$~\cite{Cleuziou2006Oct, DellaRocca2007Sep, Sochnikov2015Feb, English2016Sep, vanWoerkom2017Sep,  Spanton2017Dec, Kringhoj2018Feb, Leblanc2024May, Portoles2022Nov}.
The supercurrent of the $n^\mathrm{th}$ harmonic arises from higher-order Cooper-pair transport events, carrying a charge of $n~\times~2e$, where $e$ is the electron charge.

Josephson elements with nonsinusoidal CPR have been used to realize a variety of superconducting devices, such as parametric amplifiers~\cite{Abdo2013Jul,Sivak2020Feb,Schrade2023Oct}, supercurrent diodes~\cite{Kokkeler2022Dec, Zhang2022Nov, Souto2022Dec, Ciaccia2023Apr, Maiani2023Jun, Valentini2023Jun}, superconducting nonlinear asymmetric inductive elements (SNAILs)~\cite{Frattini2017May}, and superconducting low-inductance undulatory galvanometers (SLUGs)~\cite{Hover2012Feb}.
Moreover, circuits with tunable Josephson couplings, $E_\mathrm{J}$, and harmonic content allow for the emulation of novel quantum states of matter~\cite{Doucot2002May, Kuzmanovski2023Dec, Maffi2024May, Chirolli2024Aug} and for engineering of the eigenstates of superconducting qubits~\cite{Ye2021Jul, Gyenis2021Sep, Strickland2024Jun}, enabling the realization of parity-protected superconducting qubits with intrinsically suppressed relaxation and dephasing~\cite{Ioffe2002Dec, Brooks2013May, Bell2014Apr, Smith2020Jan, Larsen2020Jul, Gyenis2021Mar, Schrade2022Jul}. 

Nonsinusoidal Josephson elements with tunable CPRs can be constructed by combining multiple tunneling JJs whose critical currents and individual CPRs are fixed by the junction geometry~\cite{Dolan1977Sep}.
In such circuits, higher harmonics originate from internal degrees of freedom of the phase configuration~\cite{Golubov2004Apr, Barash2018Jun, Bozkurt2023Jul, Banszerus2024Feb}.

Previous work on synthesizing nonsinusoidal CPRs has primarily made use of circuits with multiple fixed SIS junctions \cite{Doucot2002May, Pop2008Sep, Gladchenko2009Jan, Bell2014Apr,Frattini2017May}.
In parallel, advances in hybrid materials have led to semiconducting JJs with intrinsically nonsinusoidal CPR~\cite{Hart2019Aug, Nichele2020Jun, Nanda2017}. The Josephson coupling and transparency of semiconductor junctions  can be tuned {\em in situ} by electrostatic gating~\cite{Deacon2017Apr, Shabani2016Apr, Kjaergaard2017Mar, Goffman2017Sep, Kayyalha2020Jan}, opening new opportunities for deterministic control over symmetries and the CPR in hybrid circuits.

In this Letter, we combine these approaches by introducing a versatile superconducting element---the hybrid Josephson rhombus---consisting of four voltage-controlled semiconducting junctions in a superconducting loop.
Comparing transport measurements to a simple model, we demonstrate that the hybrid rhombus can be tuned to realize a $\cos(2\varphi)$ element with charge-$4e$ supercurrent or a Josephson diode with nonreciprocal supercurrents. This is achieved by gate-tuning the Josephson energies, $E_{\rm J}$, of the four junctions and the magnetic frustration, $f$, of the loop; see Fig.~\ref{f1}.
The dependence of the CPR on macroscopic parameters ($E_{\rm J}$) rather than mesoscopic details (such as the number of modes and their transparencies) offers a major advantage. This approach enables deterministic tuning of the rhombus into regimes of $\cos(2\varphi)$ potentials and supercurrent diode effect.
In contrast, achieving non-sinusoidal CPRs in hybrid SQUIDs requires careful fine-tuning, limited to a convoluted control of the total current amplitude and combined mode transparencies~\cite{Hart2019Aug}.

\section{Device model}

Before presenting the experimental results, we introduce the device model and examine several characteristic configurations.
Within our model, each rhombus arm contains two JJs with sinusoidal CPRs connected in series. The CPR of each arm is given by~\cite{Bozkurt2023Jul, Banszerus2024Feb}

\begin{equation}\label{eq:cpr}
 I(\varphi) =\frac{e\sigma}{2\hbar}\frac{4\rho}{(1+\rho)^2}\frac{\sin(\varphi)}{\sqrt{1-\frac{4\rho}{(1+\rho)^2}\sin^2(\varphi/2)}}\,,
\end{equation}
where $\sigma = E_\mathrm{J1}+E_\mathrm{J2}$ is the sum of two Josephson energies in the top arm, and  $\rho=E_\mathrm{J1}/E_\mathrm{J2}$ is their ratio.
We note that Eq.~\eqref{eq:cpr} has the same form as the CPR of a single-mode junction~\cite{Beenakker1991Dec} with transparency $\tau=4\rho/(1+\rho)^2$~\cite{Bozkurt2023Jul}. The CPR is sinusoidal for $\rho \ll 1$ and $\rho \gg 1$, and highly nonsinusoidal for $\rho \approx 1$, maintaining $\tau \gtrsim 0.9$ for $0.5 < \rho < 2$.
Thus, two sinusoidal JJs in series act as a synthetic single-mode JJ, where $\sigma$ and $\rho$, and hence the amplitude and transparency of the resulting CPR, can be tuned independently by controlling $E_\mathrm{J1}$ and $E_\mathrm{J2}$. 

Interference of supercurrents through the top and bottom arms is set by magnetic frustration, $f=\Phi_\mathrm{R}/\Phi_0$, where $\Phi_\mathrm{R}$ is the magnetic flux through the rhombus and $\Phi_0 = h/2e$ is the quantum of flux.
The model CPR for the full rhombus, including four Josephson couplings along with frustration dependence, was obtained numerically by minimizing the total energy with respect to the phases on the islands between junctions in each arm, as described in Appendix~\ref{appendix:model}.

\begin{figure}[!t]
\includegraphics[draft=false,keepaspectratio=true,clip,width=\linewidth]{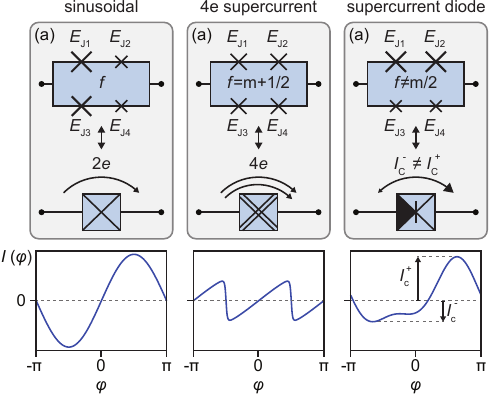}
\caption[Fig01]{
Schematic illustrations of three configurations of a Josephson rhombus.
The size of the crosses indicates the strength of the Josephson coupling, $E_\mathrm{J}$. 
(a)~For strongly asymmetric tuning within each arm, the rhombus displays a sinusoidal CPR with $f$-dependent amplitude. 
(b)~In a balanced configuration with four equal $E_\mathrm{J}$ and at full frustration ($f=m+1/2$, with integer $m$), supercurrent through the rhombus is carried by pairs of Cooper pairs (along with higher even multiples).
(c)~For symmetric tuning within each arm but strong imbalance between  arms, the rhombus exhibits a superconducting diode effect away from time-reversal symmetry preserving frustrations ($f\neq m/2$).
}
\label{f1}
\end{figure}

With $\rho$ set far from 1 in both arms, the rhombus behaves as a simple dc SQUID with a flux-dependent sinusoidal CPR carrying a charge-$2e$ supercurrent [Fig.~\ref{f1}(a)].
In contrast, when all four Josephson energies are equal, and the rhombus is tuned to full frustration ($f=m + 1/2$, where $m$ is an integer), the odd harmonics of the CPR interfere destructively, leaving only even harmonics [Fig.~\ref{f1}(b)].
This results in a $\pi$-periodic CPR with a dominant charge-$4e$ supercurrent component~\cite{Messelot2024May, Leblanc2024May}.
Such a $\cos(2\varphi)$ element preserves the parity of Cooper pairs and can be used as a building block for parity-protected qubits~\cite{Larsen2020Jul, Gyenis2021Sep}.

By symmetrizing the junctions in each arm while unbalancing the two arms, the rhombus functions as a supercurrent diode away from zero or full frustration [Fig.~\ref{f1}(c)].
The superconducting diode effect (SDE), where the critical current differs for the two current directions, $I_\mathrm{C}^+ \neq I_\mathrm{C}^-$, requires a nonsinusoidal Josephson element with broken inversion and time-reversal symmetries~\cite{Baumgartner2022Jan, Baumgartner2022Feb, Davydova2022Jun, Lin2022Oct, Hou2023Jul, Chen2024Feb, Yerin2024Apr}.
The inversion symmetry can be explicitly broken by tuning the CPR amplitudes, $\sigma$, of each arm to be different while keeping individual $\rho=1$.
To break time-reversal symmetry, the rhombus can be frustrated with a value, $f \neq m/2$.
The resulting CPR has a maximum, $I_\mathrm{C}^+$ that is different in magnitude from the minimum, $I_\mathrm{C}^-$ [Fig.~\ref{f1}(c)].

\section{Experimental setup}

\begin{figure}[!t]
\includegraphics[draft=false,keepaspectratio=true,clip,width=\linewidth]{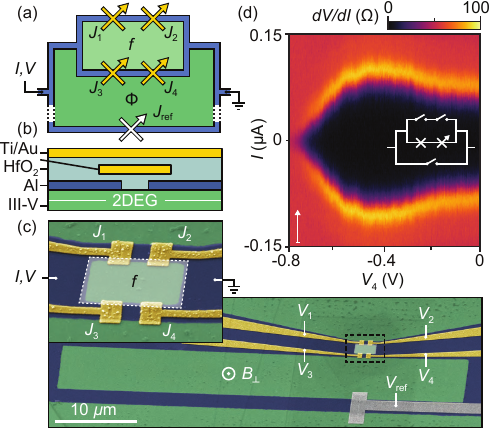}
\caption[Fig02]{ 
    (a) Schematic of the measured device comprising a hybrid Josephson rhombus with four JJs embedded in a loop with a reference JJ for phase biasing.
    (b) Schematic cross-section of a JJ illustrating the dual-gate geometry. The lower gate tunes the Josephson couplings, while the global top gate depletes the surrounding 2DEG.
    (c) Scanning electron micrograph of a reference device, taken before the deposition of the global top gate, with false color representing different materials.
    The superconducting Al is blue, and the semiconductor heterostructure is green. 
    The setup for studying transport through the rhombus is highlighted, with conducting (depleted) JJs under yellow (gray) gates.
    The inset shows a zoom-in on the rhombus. 
    (d) Differential resistance, $dV/dI$, as a function of current bias, $I$, and $J_4$ gate voltage, $V_\mathrm{4}$, with open $J_3$ ($V_\mathrm{3} = 0$) and depleted $J_\mathrm{ref}$ and $J_1$ ($V_\mathrm{1} = -1$~V, $V_{\rm ref} = - 1.5$~V), shows a nonhysteretic behavior, characteristic for an overdamped junction. The white arrow indicates the current sweep direction.
}
\label{f2}
\end{figure}

The hybrid rhombus studied here is based on a two-dimensional electron gas (2DEG) of InAs, proximitized by an epitaxial Al film~\cite{Shabani2016Apr}, using the same approach as in Ref.~\cite{Banszerus2024Feb}.
The superconducting Al loop with four JJs, denoted $J_1$ to $J_4$, was patterned by selective Al etching.
The hybrid rhombus was embedded in a larger loop with a reference junction, $J_\mathrm{ref}$, designed to support a higher critical current compared to the rhombus [Fig.~\ref{f2}(a)].
Each JJ was equipped with an individual Ti/Au top gate after depositing a thin HfOx dielectric layer, allowing for independent control of all five JJs.
A second layer of HfOx was deposited before metalizing a global Ti/Au gate to deplete the surrounding 2DEG.
A schematic cross-section through one of the JJs is illustrated in Fig.~\ref{f2}(b).
A scanning electron micrograph of a reference device without the top gate is shown in Fig.~\ref{f2}(c).
Two devices showing similar behavior have been measured.
The differential resistance, $dV/dI$, was measured in a dilution refrigerator with a three-axis vector magnet and base temperature of 20~mK using standard ac lock-in techniques in a four-terminal configuration.
Further details on wafer structure, sample fabrication, and measurements are given in Appendix~\ref{appendix:fabrication}.

Before exploring the different regimes of the rhombus, we characterize the gate-voltage dependence of the supercurrent transport through each junction.
A representative measurement of $dV/dI$ as a function of current bias, $I$, and gate voltage $V_4$, taken while keeping $J_\mathrm{1}$, $J_\mathrm{2}$, and $J_\mathrm{ref}$ pinched-off and $J_\mathrm{3}$ fully open, is show in Fig.~\ref{f2}(d).
As the gate voltage is decreased, the switching current, $I_{\rm SW}$, first increases to about 100~nA at $V_\mathrm{4} \sim -0.5~V$, before getting suppressed around $-0.8~V$. The transitions between the superconducting and finite resistance regions in the top (bottom) half of the map mark the switching (retrapping) current of the JJ. The high symmetry between switching and retrapping currents suggests that the junctions are overdamped.
We use these data to estimate the four gate-dependent Josephson energies of the rhombus, $E_\mathrm{J1}(V_1)$ to $E_\mathrm{J4}(V_4)$, which are then used for device modeling. 

\begin{figure}[!t]
\includegraphics[draft=false,keepaspectratio=true,clip,width=\linewidth]{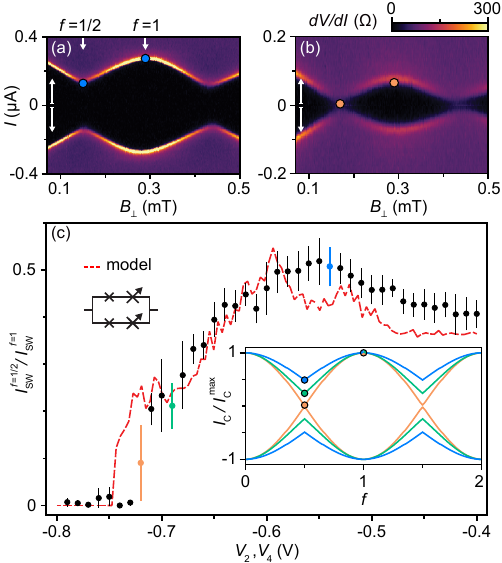}
\caption[Fig03]{
    (a) Differential resistance, $dV/dI$, as a function of current bias, $I$, and perpendicular magnetic field, $B_\perp$, measured for the hybrid rhombus with the gate voltages tuned such that the four JJs have matching switching currents, $(V_1, V_2, V_3, V_4) = (-0.54, -0.63, 0, -0.63)$~V.
    The blue points mark $I_\mathrm{SW}$ at full ($f=1/2)$ and integer ($f=1$) frustration.
    The white arrows indicate the current sweep direction.
    (b) Same as (a) but with strongly asymmetric tuning of switching currents within each arm, $(V_1, V_2, V_3, V_4) = (-0.54, -0.74, 0, -0.74)$~V.
    (c) Ratio of switching currents at $f=1/2$ and $f=1$, $I_\mathrm{SW}^{f=1/2}/I_\mathrm{SW}^{f=1}$, as a function of $V_2$ and $V_4$, simultaneously varied between $-0.8$ and $-0.4$~V, while keeping $V_1=-0.54$~V and $V_3=0$ fixed.
    The dashed red curve is the expected $I_\mathrm{SW}^{f=1/2}/I_\mathrm{SW}^{f=1}$, based on a numerical model using the measured $I_\mathrm{SW}$ of the individual junctions.
    Inset: calculated $I_\mathrm{SW}$ as a function of $f$ for different symmetrization tunings.
}
\label{f3}
\end{figure}

\section{Nonsinusoidal CPR and charge-4\MakeLowercase{\textit{e}} supercurrent}

\begin{figure}[!t]
\includegraphics[draft=false,keepaspectratio=true,clip,width=\linewidth]{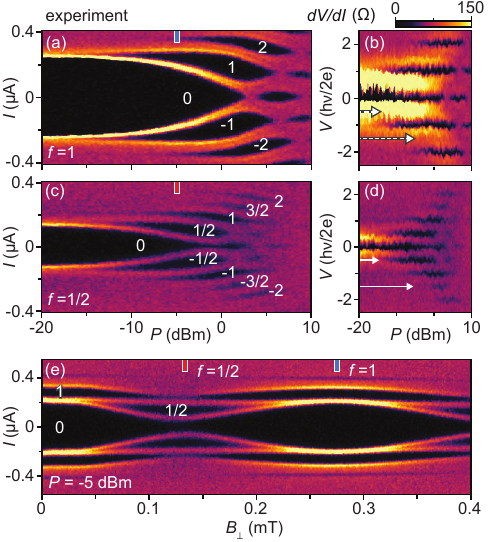}
\caption[Fig04]{
(a) Measured differential resistance, $dV/dI$, as a function of dc current, $I$, and applied rf radiation power, $P$, taken for rhombus at integer frustration, $f=1$, and a fixed rf frequency of $\nu=4$~GHz.
The data show Shapiro lobes forming as $P$ is increased.
(b) Parametric plot of $dV/dI$ from (a) as a function of simultaneously measured dc voltage across the rhombus, $V$, and $P$.
Shapiro steps occur at integer values of $V = h\nu/2e$, indicating predominantly charge-$2e$ transport. 
(c)--(d) Same as (a)--(b) but at full frustration, $f=1/2$, with Shapiro steps occurring at half-integer values of $V = h\nu/2e$, in agreement with dominant charge-$4e$ supercurrent.
(e) Measured $dV/dI$ as a function of $I$ and perpendicular magnetic field, $B_\perp$, taken at $\nu=4$~GHz and $P=-5$~dBm.
}
\label{f4}
\end{figure}

\begin{figure}[!t]
\includegraphics[draft=false,keepaspectratio=true,clip,width=\linewidth]{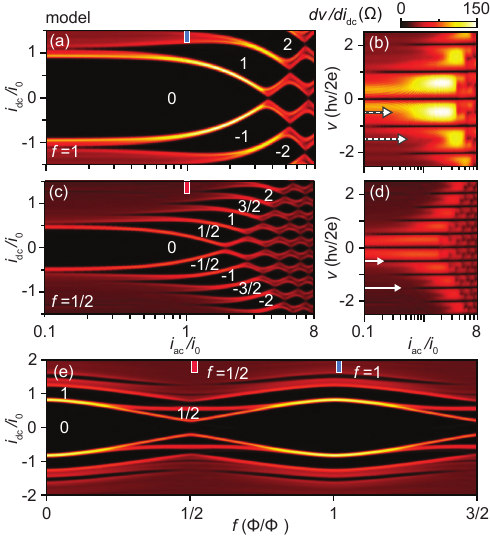}
\caption[Fig05]{
(a) Calculated differential resistance, $dv/di_\mathrm{dc}$, as a function of dc and ac currents, $i_\mathrm{dc}$ and $i_\mathrm{ac}$, modeled for a rhombus at integer frustration, $f=1$. 
(b) Parametric plot of $dv/di_\mathrm{dc}$ from (a) as a function of the dc voltage, $v$, and $i_\mathrm{ac}$.
Shapiro steps occur at multiples of $v = h\nu/2e$.
\mbox{(c)--(d)}~Same as (a)--(b) but at full frustration, $f=1/2$, with Shapiro steps occurring at multiples of $v = h\nu/4e$.
Agreement with experimental data in Fig.~\ref{f4} supports the interpretation of dominant charge-$4e$ transport at $f=1/2$.
(e) Calculated $dv/di_\mathrm{dc}$ as a function of $i_\mathrm{dc}$ and $f$, modeled for an ac current $i_\mathrm{ac}/i_0 = 1$, where $i_0$ is the critical current at $i_\mathrm{ac}=0$.
Calculations were done using Stewart-McCumber parameter $\beta=0.1$ and reduced frequency $\Omega=1.2$; see Appendix~\ref{appendix:model}.
}
\label{f5}
\end{figure}

We begin by experimentally investigating junction asymmetry (effective transparency) of both arms with the reference junction depleted.
The differential resistance, $dV/dI$, as a function of current bias, $I$, and perpendicular magnetic field, $B_\perp$, taken with all four junctions fully balanced shows an oscillatory $I_\mathrm{SW}$ that remains finite throughout the measured range; see Fig.~\ref{f3}(a).
The switching current is maximal at $B_\perp\approx0.28$~mT, corresponding to one flux quantum threading the rhombus loop ($f=1$), and minimal but finite---roughly half of the total amplitude---at $B_\perp\approx0.14$~mT ($f=1/2$).
This contrasts the imbalanced configuration with $\rho \ll 1$ in both arms, where the supercurrent is suppressed at $f=1/2$, resembling a symmetric dc SQUID; see Fig.~\ref{f3}(b).

To further investigate the origin of the finite supercurrent at $f=1/2$, we measure the ratio of the switching currents at $f=1/2$ and $f=1$, $I_\mathrm{SW}^{f=1/2}/I_\mathrm{SW}^{f=1}$, as a function of $V_2$ and $V_4$, while keeping $V_1$ and $V_3$ fixed; see Fig.~\ref{f3}(c).
Thanks to the relaxed balancing condition for $\rho$, we use a common gate voltage for two JJs being tuned. 
The ratio peaks around 0.5 at $V_2 = V_4 \approx -0.55$~V, near the balancing point of the rhombus, and decays when the voltages are detuned away from this configuration.

We use the measured gate-voltage dependence of the individual switching currents of the four JJs to model the rhombus CPR, from which we calculate $I_\mathrm{SW}^{f=1/2}/I_\mathrm{SW}^{f=1}$ as a function of $V_2$ and $V_4$; see dashed red curve and inset in Fig.~\ref{f3}(c).
The model reproduces $I_\mathrm{SW}^{f=1/2}/I_\mathrm{SW}^{f=1}\approx0.5$ at full balancing and the suppression of $I_\mathrm{SW}$ at $f=1/2$ for strong asymmetry within the arms.
According to the model, the remaining supercurrent at $f=1/2$ is due to even harmonics, which interfere constructively at full frustration.
This suggests that the finite supercurrent at $f=1/2$ originates from a dominant $\cos(2\varphi)$ component.

The dominant Josephson harmonic that determines the charge character (2e versus 4e) of the supercurrent can be probed explicitly by studying the dc response of the rhombus to radio frequency (rf) radiation~\cite{Shapiro1963Jul, Wiedenmann2016Jan, Bocquillon2017Feb, Ueda2020Sep, Ciaccia2023Jun, Huang2023Jun}.
The differential resistance, $dV/dI$, as a function of current bias, $I$, and rf power, $P$, measured at $f=1$ and rf frequency $\nu =4$~GHz in a balanced configuration displays well-defined Shapiro lobes with vanishing resistance, separated by $dV/dI$ peaks; see Fig.~\ref{f4}(a).
The accompanying measurements of dc voltage, $V$, show Shapiro steps developing at multiples of $V=h\nu/2e$; see Appendix~\ref{appendix:measurements}.
We find that plotting $dV/dI$ as a function of $P$ and $V$ parametrically is a good way to illustrate the nature of the dominant charge character reflected in the Shapiro step heights [Fig.~\ref{f4}(b)].
Specifically, the dips in $dV/dI$ appear at integer values of $h\nu/2e$, as expected for predominantly charge-$2e$ supercurrent at integer frustration~\cite{Wiedenmann2016Jan, Ueda2020Sep, Ciaccia2023Jun}. 

Repeating the measurement at $f=1/2$, we observe a higher number of the Shapiro lobes [Fig.~\ref{f4}(c)], with steps occurring at half-integer multiples of $h\nu/2e$~[Fig.~\ref{f4}(d)].
The emergence of the intermediate Shapiro steps indicates a transition from a charge-$2e$ to charge-$4e$ dominated supercurrent at full frustration~\cite{Ciaccia2023Jun,Valentini2023Jun}.
The frustration-driven transition can be clearly followed in a $dV/dI$ map as a function of $I$ and $B_\perp$ taken at a fixed rf excitation of \mbox{$P=-5$}~dBm [see Fig.~\ref{f4}(e)]. The data shows the emergence of a clearly visible half-integer Shapiro step (\mbox{$V=h\nu/4e$}) around $B_\perp=0.14$~mT, corresponding to $f=1/2$.
We note that the monotonous decrease in the size of the Shapiro lobes suggests a strong suppression of the $\cos(\varphi)$ harmonic and a dominating $\cos(2\varphi)$ harmonic. A finite  $\cos(\varphi)$ term would instead result in an odd-even sequence of the size of the observed Shapiro lobes~\cite{Ueda2020Sep}, as observed for the case of imbalanced rhombus arms; see Appendix~\ref{appendix:RF_unb}.

To support our experimental findings, we numerically calculate the Shapiro response using the RCSJ model~\cite{Stewart1968Apr, McCumber1968Jun, Park2021Jun}, in which the hybrid rhombus is represented as a parallel circuit of a lumped Josephson element, a resistor, and a capacitor.
We obtain the CPR of the Josephson element using the classical model of the rhombus.
This CPR is then used to solve the RCSJ equation of motion in $\varphi$-space, yielding the current-voltage characteristics of the rhombus; see Appendix~\ref{appendix:rf} for further details on the modeling of the rf response.
The calculated rf response reproduces the main experimental observations shown in Fig.~\ref{f4}, particularly the emergence of half-integer Shapiro steps around $f=1/2$, originating from a $\pi$-periodic CPR.
The agreement between experimental data and the model suggests that the rhombus remains in its ground state, and the rf drive can be considered adiabatic.
In this case, the rhombus can be treated as a single Josephson element without additional internal degrees of freedom.
We expect this approximation to hold for rf driving $\nu < 2eI_\mathrm{C} R_\mathrm{N}/\hbar\approx 20~$GHz~\cite{SeoaneSouto2024Apr}, with the normal state resistance, $R_\mathrm{N}$, making the hybrid rhombus compatible with typical frequencies used in circuit quantum electrodynamics~\cite{Larsen2020Jul}. 

\begin{figure}[!t]
\includegraphics[draft=false,keepaspectratio=true,clip,width=\linewidth]{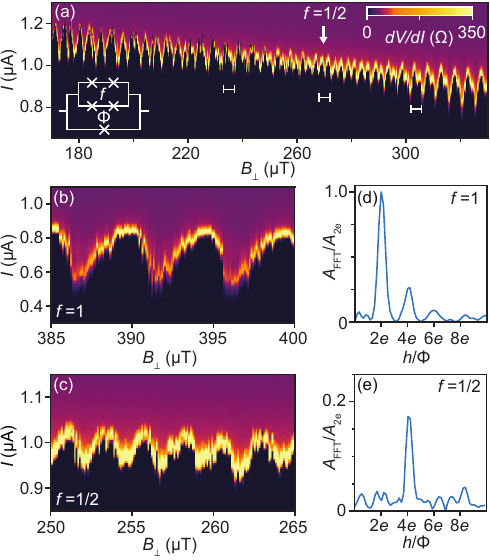}
\caption[Fig06]{
(a) Differential resistance, $dV/dI$, as a function of current bias, $I$, and perpendicular magnetic field, $B_\perp$, measured with $J_{\rm ref}$ open ($V_\mathrm{ref}=0$) in a balanced rhombus configuration.
The field range is shifted with respect to the previous figures due to hysteresis of the superconducting magnet.
(b)--(c) Similar to (a) but over a smaller $B_\perp$ range around (b) $f=1$ and (c) $f=1/2$.
(d) Fast Fourier transform amplitude of $\Delta I_\mathrm{SW}$ from (b), in the range of $B_\perp=385$ to $415~\mu$T.
(e) Same as (d) but for data from (c), in the range of $B_\perp=235$ to $265~\mu$T.
}
\label{f6}
\end{figure}

For a more direct measure of the CPR, we open $J_{\rm ref}$ by setting $V_\mathrm{ref} = 0$ and phase bias the rhombus by applying $B_\perp$~\cite{Murani2017Jul, Nichele2020Jun}. 
The much larger reference loop area allows us to phase bias the rhombus through multiple CPR periods while keeping $f$ nearly constant.
In this configuration, $dV/dI$ as a function of $I$ and $B_\perp$ displays $I_\mathrm{SW}\approx1~\mu$A, modulated by nonsinusoidal oscillations with field-dependent frequency and amplitude; see Fig.~\ref{f6}(a).
We attribute the large background component of $I_\mathrm{SW}$ to $J_\mathrm{ref}$ and the oscillatory component, $\Delta I_\mathrm{SW}$, to the flux-induced modulation of phase difference across the rhombus.
The data displays several discontinuous, non-repeatable jumps, which we associate with phase slips likely originating from the large inductance of the reference loop.
When tuning the frustration toward $f=1/2$, we first observe the appearance of a beating pattern in oscillations of $\Delta I_\mathrm{SW}$ and finally a halving of the periodicity in a narrow interval around $f = 1/2$.
The change in the periodicity can be better seen in the zoom-in maps around $f=1$ and $f=1/2$ [Figs.~\ref{f6}(b) and \ref{f6}(c)].
The data taken around $f=1$ reveals nonsinusoidal $I_{\rm SW}$ oscillations with a period of $\Delta B_\perp\approx 4.8~\mu$T and $\Delta I_\mathrm{SW} \approx 200$~nA.
At full frustration ($f=1/2$), the modulation frequency doubles while the amplitude halves.
A fast Fourier transform of $\Delta I_\mathrm{SW}$ measured at $f=1$ displays considerable contributions to the CPR from the first three harmonics ($2e, 4e$, and $6e$); see Fig.~\ref{f6}(d).
This is consistent with the nonsinusoidal nature of the two balanced JJs in series in each arm of the rhombus~\cite{Bozkurt2023Jul, Bozkurt2023Sep, Banszerus2024Feb}.
At $f=1/2$, the Fourier spectrum reveals that the CPR is dominated by the second harmonic, while the amplitudes of the first and third harmonics are suppressed.
The observed change in periodicity can be understood by the destructive interference of the odd harmonics at $f=1/2$, resulting in a dominant $\cos(2\varphi)$ term, where the supercurrent is predominantly carried by tunneling events of pairs of Cooper pairs~\cite{Doucot2002May, Pop2008Sep, Gladchenko2009Jan,Giuliano2010Oct, Ciaccia2023Jun}.
At $f=1/2$, $A_\mathrm{2e}$ is indistinguishable from the noise floor of the Fourier spectrum, indicating that the first harmonic is suppressed by more than 96\%, compared to integer frustration.
The stochastic switching events are uncorrelated with $B_\perp$ and thus do not result in specific frequency contributions in Fig.~\ref{f6}(d) and \ref{f6}(e).

We note that a finite inductance of the loop may lead to self-biasing and distortion of the CPR.
Following Ref.~\cite{Annunziata2010Oct}, we estimate the kinetic inductance of the entire device to be around $L_\mathrm{K} = 300$~pH.
The geometric inductance, $L_\mathrm{G}\approx 40$~pH, is negligible in comparison.
This indicates that inductive self-biasing can cause nonlinearities between the total and external fluxes, $\Phi=\Phi_\mathrm{ext}-L_\mathrm{K} \Delta I_{\rm SW}(\varphi)$, up to roughly $2\%$ of $\Phi_0$ at integer frustration~\cite{Nichele2020Jun}.
The effect is suppressed further around $f=1/2$ where $\Delta I_\mathrm{SW}$ is reduced.
Furthermore, it has recently been reported that accurate measurements of the CPR in a SQUID geometry require a sufficiently high asymmetry in magnitudes of derivatives of the supercurrents in the two arms~\cite{Babich2023Jul}.
Our numerical simulations using experimentally determined device parameters show that the observed $I_{\rm SW}$ oscillations accurately represent the rhombus CPR.



\section{Superconducting diode effect}

\begin{figure}[!t]
\includegraphics[draft=false,keepaspectratio=true,clip,width=\linewidth]{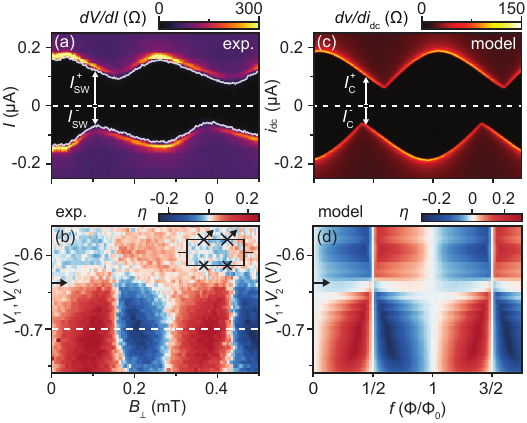}
\caption[Fig07]{
(a)~Measured differential resistance, $dV/dI$, as a function of current bias, $I$, and perpendicular magnetic field, $B_\perp$, taken for the rhombus with the gate voltages tuned such that the JJs within each arm are balanced, but strongly imbalanced between the two arms $(V_1, V_2, V_3, V_4) = (-0.70, -0.70, 0, -0.54)$~V.
The solid white curves indicate the extracted switching currents, $I_\mathrm{SW}^+$ and $I_\mathrm{SW}^-$, measured by sweeping $I$ away from zero.
(b)~Measured diode efficiency, $\eta$, as a function of $V_{1}$ and $V_2$ varied simultaneously, and $B_\perp$, taken while keeping $V_{3}=0$ and $V_{4}=-0.54$~V fixed. 
(c)~Calculated differential resistance, $dv/di_\mathrm{dc}$, as a function of dc current, $i_\mathrm{dc}$, and $f$. 
(d)~Modeled $\eta$ as a function of $V_{1}$ and $V_2$, and $f$.
Calculations were done using the gate-dependent switching currents of the individual JJs as input; see Appendix \ref{appendix:model}. 
}
\label{f7}
\end{figure}

Finally, we demonstrate that the hybrid rhombus can be tuned to a regime exhibiting SDE, expected to emerge in nonsinusoidal Josephson elements with broken inversion and time-reversal symmetries~\cite{Baumgartner2022Jan, Baumgartner2022Feb, Davydova2022Jun, Lin2022Oct, Hou2023Jul, Chen2024Feb}.
To achieve this, we symmetrize the two junctions in each arm, while tuning the balance between the two arms.
This configuration explicitly breaks inversion symmetry due to different CPR amplitudes, $\sigma$, in each arm while maintaining the higher harmonics, thanks to the individual $\rho \approx 1$.
Time-reversal symmetry is explicitly broken by frustrating the rhombus.

The measured $dV/dI$ displays a periodic switching current whose amplitude at a given $B_\perp$ depends on the direction of the supercurrent, $I_\mathrm{SW}^+ \neq I_\mathrm{SW}^-$ [white arrows in Fig.~\ref{f7}(a)].
To quantify the strength of the effect, we study the diode efficiency, $\eta = (I_\mathrm{SW}^+ - I_\mathrm{SW}^-)/(I_\mathrm{SW}^+ + I_\mathrm{SW}^-)$, as a function of $B_\perp$ and combined gate voltages $V_1=V_2$, which control the supercurrent amplitude in one of the arms and thus the balancing of the rhombus; see Fig.~\ref{f7}(b).
The observed $\eta$ oscillates with the field, changing its sign around frustrations that conserve time-reversal symmetry ($f=1/2, 1$, and $3/2$) and reaching values close to $\pm25\%$ in between.
Furthermore, $\eta$ exhibits an additional sign flip around $V_1 = V_2 \approx -0.64$~V, at the balancing point of the two rhombus arms, where inversion symmetry is~present.

We compare the measured data with the numerical model of the device discussed in Appendix~\ref{appendix:model}.
The calculated differential resistance, $dv/di_\mathrm{dc}$, as a function of the dc current, $i_\mathrm{dc}$, and frustration, $f$, for the same gate configuration as in Fig.~\ref{f7}(a) displays nonreciprocal critical current, $I_\mathrm{C}^+ \neq I_\mathrm{C}^-$  [Fig.~\ref{f7}(c)].
The modeled diode efficiency, $\eta = (I_\mathrm{C}^+ - I_\mathrm{C}^-)/(I_\mathrm{C}^+ + I_\mathrm{C}^-)$, agrees well with the directly measured values of $\eta$ and reproduces the vanishing $\eta$ for configurations with conserved time-reversal ($f=1/2, 1$, and $3/2$) and inversion ($V_1 = V_2 = -0.64$~V) symmetries. 

\section{Discussion}
The realization of semiconductor-superconductor hybrid materials opens up new opportunities for the next-generation quantum devices with \textit{in situ} gate-voltage control.
In this work, we have introduced a hybrid Josephson rhombus consisting of four Josephson junctions in a loop, defined in an epitaxial InAs/Al heterostructure.
By constructing superconducting circuits consisting of multiple hybrid junctions, our approach combines the advantages of voltage control from SNS systems and the deterministic behavior of SIS circuits, making the rhombus resilient against microscopic device-to-device variations.
The harmonic content of each rhombus arm can be tuned from sinusoidal to nonsinusoidal by symmetrizing the switching currents of the junctions~\cite{Bozkurt2023Jul, Banszerus2024Feb}.
The current phase relation of the rhombus can be further tailored by introducing frustration to the superconducting loop.
Using these experimental knobs, we have demonstrated that the hybrid rhombus can be tuned to either a balanced $\cos(2\varphi)$ element that preserves Cooper-pair parity or a superconducting diode with highly nonreciprocal supercurrent transport.

We anticipate that the ability to balance the hybrid Josephson rhombus \textit{in situ} will enable novel voltage-controlled qubit operation schemes and be beneficial for the realization of parity-protected qubits with improved coherence, which would decohere more rapidly in the presence of finite $\cos(\varphi)$ component~\cite{Smith2020Jan, Schrade2022Jul}.
Furthermore, the independent voltage control of the amplitude and harmonic content of the hybrid rhombus CPR holds promise for tunable parametric amplifiers~\cite{Abdo2013Jul, Sivak2020Feb}.
The ability to synthesize and suppress harmonics of the CPR, combined with the simultaneous control over time reversal and inversion symmetry, will be particularly useful in complex devices, such as the Josephson synthesizer, which enable the generation of arbitrary current phase relations~\cite{Bozkurt2023Jul}.
The deterministic dependence of the CPR on macroscopic JJ parameters rather than on mesoscopic details offers a major advantage over relying on the non-sinusoidal CPR of high-transparency hybrid JJs, particularly for scalable technological applications.

\begin{acknowledgments} 
We thank A.~C.~C.~Drachmann for assistance with device fabrication and M. Kjaergaard for discussions. We acknowledge support from research grants (Projects No. 43951 and No. 53097) from VILLUM FONDEN, the Danish National Research Foundation, and the European Research Council (Grant Agreement No. 856526).
\end{acknowledgments} 

\revappendix

\section{MODELLING OF THE RHOMBUS} 
\label{appendix:model}
\label{appendix:rf}

\subsection{CURRENT PHASE RELATION} 

We model the current phase relation of the device by considering four sinusoidal JJs arranged in a superconducting loop as shown in Fig.~\ref{f1}.
The total energy, $E_\mathrm{tot}$, of the superconducting circuit in the classical limit, with vanishing charging energy, is given by the sum over the four Josephson energies,
\begin{equation*}
    E_\mathrm{tot} = \sum_{i=1}^4 E_{\mathrm{J}i}(1-\cos\chi_i)\,,
\end{equation*}
where $\chi_\mathrm{i}$ is the superconducting phase difference across the $i^{th}$ JJ.
To find the ground state energy, $E_\mathrm{0}(f, \varphi)$, we numerically minimize $E_\mathrm{tot}$ with respect to $\chi_\mathrm{i}$ under the following boundary conditions imposed by the frustration, $f$, and by the phase bias, $\varphi$, across the rhombus,
\begin{equation*}
\sum_\mathrm{i=1}^4 \chi_i = 2 \pi f\,, \qquad \chi_1 + \chi_2 =\pi f + \varphi\,.
\end{equation*}
Knowing the ground state energy, we can calculate the CPR as

\begin{equation}\label{eq:derive_cpr}
 I(f, \varphi) =\frac{2e}{\hbar}\frac{\partial E_\mathrm{0} (f, \varphi)}{\partial \varphi}\,.
\end{equation}
The positive (negative) critical current of the rhombus, $I_\mathrm{C}^{+}$ ($I_\mathrm{C}^{-}$), is given by the maximum (minimum) of the CPR for a given $f$.
Depending on the functional form of $I(f, \varphi)$, the magnitudes of $I_\mathrm{C}^{+}$ and $I_\mathrm{C}^{-}$ may be different, signifying the superconducting diode effect.

Our model assumes that the four JJs in the rhombus have sinusoidal CPRs.
To test the validity of this simplification, we numerically calculate the CPR of two transparent single-mode JJs in series, representing a single arm of the hybrid rhombus.
The intrinsic transparencies of the modes, $\tau_\mathrm{int}$, determine the harmonic content of each junction CPR, given by~\cite{Beenakker1991Dec}  
\begin{equation*}
    I(\varphi)=\frac{e\Delta}{2\hbar}\frac{\tau_\mathrm{int}\,\sin(\varphi)}{\sqrt{1-\tau_\mathrm{int}\,\sin^2(\varphi/2)}}.
\end{equation*}

We find that sinusoidal CPRs remain a good approximation as long as the effective transparency of a rhombus arm, $\tau=4\rho/(1+\rho)^2$, remains larger than individual $\tau_\mathrm{int}$~\cite{Banszerus2024Feb}.
In practice, this implies that the intrinsic JJ transparencies limit the extent to which the rhombus can be tuned to behave as a purely sinusoidal Josephson element; see Fig.~\ref{f1}(a).
The simplified model reproduces the main features of the experiment across different tuning regimes, which further supports the validity of this approximation.

\subsection{DIODE EFFECT}
\begin{figure}[!b]
    \centering
    \includegraphics[width=1\linewidth]{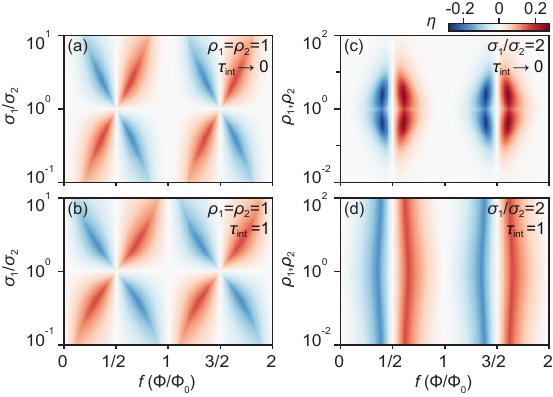}
    \caption{
    (a) Modelled diode efficiency, $\eta$, as a function of $\sigma_1/\sigma_2$ and $f$ for a hybrid rhombus containing 4 sinusoidal JJs. 
    (b) Same as (a) but for a hybrid rhombus containing 4 fully transparent, single-mode JJs. 
    (c) Modelled diode efficiency, $\eta$, as a function of $\rho_1=\rho_2$ and $f$ for a hybrid rhombus containing 4 sinusoidal JJs. 
    (d) Same as (c) but for a hybrid rhombus containing 4 fully transparent, single-mode JJs. 
     }
    \label{fa1}
\end{figure}

In this section, we study the modeled diode efficiency as a function of $\sigma$, $\rho$, and $f$ for rhombi composed of four junctions with different intrinsic junction transparencies, $\tau_{\rm int}$.
Specifically, we compare the case of a rhombus with four sinusoidal ($\tau_{\rm int} \rightarrow 0$)  JJs and a rhombus with four fully transparent ($\tau_{\rm int} = 1$) JJs; see Fig.~\ref{fa1}. 
We first investigate the diode efficiency, $\eta$, as a function of $\sigma_1/\sigma_2$ and $f$ for a hybrid rhombus in two limiting cases [Fig.~\ref{fa1}(a) and \ref{fa1}(b)].
We note that in the limit of $\rho \approx 1$ the CPR of two JJs in series is very weakly dependent on $\tau_{\rm int}$, with only a minute increase of the amplitudes of higher harmonics for $\tau_{\rm int} =1$~\cite{Banszerus2024Feb}.
This merely results in a slight shift of the frustration value that yields the highest diode efficiencies.
Next, we investigate $\eta$, as a function of $\rho_{1} = \rho_2$ and $f$ [Fig.~\ref{fa1}(c) and \ref{fa1}(d)]. 
A rhombus consisting of four fully transparent, single-mode JJs lacks the previously observed dependence on $\rho_{1}$ and $\rho_{2}$, as the CPR remains non-sinusoidal even for $\rho_{1} = \rho_2 \gg 1$, owing to the CPR of the individual JJs.
Interestingly, the model implies that the diode efficiency for sinusoidal JJs can be further increased by tuning $\rho_\mathrm{1} = \rho_2$ away from 1.
Currently, we are unaware of a closed analytical expression that would allow estimating the maximal diode efficiency in the phase space of $f, \sigma_{1}, \sigma_{2}, \rho_{1},$ and $\rho_{2}$.

\subsection{RF RESPONSE}

We model the response of the current-voltage characteristic to rf irradiation using the RCSJ framework~\cite{Stewart1968Apr, McCumber1968Jun, Park2021Jun}.
The rhombus is treated as a parallel circuit consisting of a lumped Josephson element with a characteristic CPR determined by Eq.~\eqref{eq:derive_cpr}, a resistor with resistance $R_\mathrm{N}$, and a capacitor with capacitance $C$.
To obtain the current-voltage characteristics, we consider a system driven by an external dc current, $I_\mathrm{dc}$, and the rf-induced ac current, $I_\mathrm{ac}$, with the equation of motion given by~\cite{Park2021Jun}  
\begin{equation}
    i_\mathrm{dc}+i_\mathrm{ac} \cos(\Omega \tau) = I[\varphi(\tau)] + \frac{\partial \varphi}{\partial \tau} + \beta \frac{\partial^2 \varphi}{\partial \tau^2}\,, 
    \label{RCSJ}
\end{equation}
where $i_\mathrm{dc\,(ac)}=I_\mathrm{dc\,(ac)}/I_\mathrm{C}$ are the dimensionless currents, $\Omega=h \nu/(2eI_\mathrm{C}R_\mathrm{N})$ is the reduced frequency, ${\tau=2eI_\mathrm{C}R_\mathrm{N}t/\hbar}$ is the reduced time, and $\beta=2eI_\mathrm{C}R_\mathrm{N}^2C/\hbar$ is the Stewart-McCumber parameter.
We solve Eq.~\eqref{RCSJ} for $\varphi(\tau)$ numerically over the interval $\tau=0$ to $\tau=40\,\pi$ for each combination of $i_\mathrm{dc}$ and $i_\mathrm{ac}$, while keeping $\beta=0.1$ and $\Omega=1.2$ constant.
The numerical values of $\beta$ and $\Omega$ were chosen to best reproduce the observed Shapiro patterns and are consistent with the experimental device parameters for $I_\mathrm{0}$ and $R_\mathrm{N}$. 
We note that $\beta=0.1$ corresponds to a capacitance of $C=22.4$~fF, which is comparable with a simple plate capacitor approximation of the gate capacitance yielding $C=13.5$~fF.   
We chose $\varphi(0)=0$ and $\partial \varphi/\partial \tau (0)=0$ as initial conditions.
The dc voltage across the rhombus, $v$, is calculated by taking the time average of $\partial \varphi/\partial \tau$ between $\tau=6\,\pi$ and $\tau=40\,\pi$, chosen to ensure that the system has reached its equilibrium state.
The differential resistance, $dv/di_\mathrm{dc}$, is obtained by numerical differentiation of $v(i_\mathrm{ac}, i_\mathrm{dc})$. 
 
By modeling the device as a single Josephson element, we implicitly assume that the rhombus remains in its ground state, $E_0$, and we disregard potential excitations of internal degrees of freedom.
We expect non-adiabatic behavior to arise in the fast-driving regime, where $\Omega~\gg~1$~\cite{SeoaneSouto2024Apr}.
Our calculations agree well with the experimental data, suggesting that this approximation is valid within the experimentally probed frequency range, $\nu=1-8$~GHz.
This range is typical for superconducting circuits, indicating that hybrid rhombi are compatible with these applications.

\begin{figure}[b]
    \includegraphics[width=\linewidth]{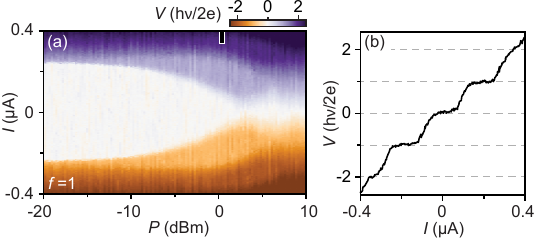}
    \caption{ 
    (a) Measured dc voltage, $V$, as a function of current bias, $I$, and applied rf radiation power, $P$, taken for rhombus at integer frustration, $f=1$, and a fixed rf frequency of $\nu=4$~GHz.
    The measurements were taken simultaneously with the data in Fig.~\ref{f4}(a).
    (b) Line-cut taken from (a) at $P=0.5$~dBm shows Shapiro steps developing at multiples of $V = h\nu/2e$.
    }
    \label{fa2}
\end{figure}

\section{SAMPLE PREPARATION}\label{appendix:fabrication}

The III-V semiconductor heterostructure was grown on a semi-insulating InP(100) substrate by molecular beam epitaxy (MBE).
It consists of a 7~nm thick InAs quantum well, encapsulated between a 4~nm thick In$_{0.75}$Ga$_{0.25}$As bottom barrier and a 10~nm thick In$_{0.75}$Ga$_{0.25}$As top barrier, followed by two monolayers of GaAs capping.
A 5~nm thick Al film was epitaxially grown \textit{in situ} without breaking the vacuum. 

The devices were patterned using electron beam lithography (Elionix, 100~keV).
The Al film was selectively etched using Transene Aluminum etch type~D at 50\,$^\circ$C for 5 seconds.
The mesa structures were defined by a chemical wet etch using (220:55:3:3 H$_2$O:C$_6$H$_8$O$_7$:H$_3$PO$_4$:H$_2$O$_2$).
The HfOx (18~nm) gate dielectrics were grown using atomic layer deposition (Veeco Savannah).
The first Ti/Au (3/20 nm) and the second Ti/Au (5/350~nm) gate layers were deposited using electron beam evaporation (AJA International) at a base pressure of $10^{-8}$~mbar.
The gate electrode above $J_3$ showed no response, presumably due to unsuccessful fabrication, and was kept at $V_3 = 0$.
The global top gate was kept at $-1.3$~V throughout the experiment.
We observed a small hysteresis of the superconducting magnet around $f=0$ and chose to focus on $f=1$ to represent integer frustration.

The InAs quantum well was characterized in a gate-controlled Hall bar device, where the epitaxial Al film was removed.
The maximum observed charge carrier mobility of $\mu=53,000$~cm$^{2}$/Vs was measured at a charge carrier density of $n=0.6\times 10^{12}~$cm$^{-2}$/Vs.
The normal state sheet resistance, $R_{\square} = 6.5~\Omega$, used for estimating the kinetic inductance, was measured separately above the critical temperature of the Al film in a Hall bar geometry.
An induced superconducting gap of $\Delta=180~\mu$eV was measured using a tunneling spectroscopy device.

\section{EXPERIMENTAL TECHNIQUES}
\label{appendix:measurements}
\label{appendix:RF_unb}

\begin{figure}[!t]
    \includegraphics[width=\linewidth]{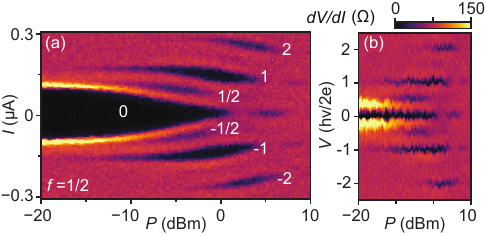}
    \caption{
    (a) Measured differential resistance, $dV/dI$, as a function of dc current, $I$, and applied rf radiation power, $P$, taken for rhombus at a frustration, $f = 1/2$, with imbalanced arms ($\sigma_1 \neq \sigma_2$), and a fixed rf frequency of $\nu = 4$ GHz. The data show an even-odd lobe pattern, with integer Shapiro lobes being more pronounced than half-integer lobes. (b) Parametric plot of $dV/dI$ from (a) as a function of simultaneously measured dc voltage across the rhombus, $V$, and $P$. Shapiro steps are well pronounced at integer values of $V= h\nu/2e$ and less pronounced at half-integer values, indicating a mix of charge-$2e$ and charge-$4e$ transport with comparable amplitudes.
     }
    \label{fa3}
\end{figure}

\subsection{MEASUREMENT SETUP}
Electrical measurements were carried out in a commercial cryofree dilution refrigerator (Oxford Instruments, Triton 400) at a base temperature of 20~mK.
Standard dc and low-frequency ac ($f=27.4$~Hz) lock-in measurement techniques were used.
All dc lines were filtered at cryogenic temperatures using a commercial two-stage RC and LC filter with a cut-off frequency of 65~kHz (QDevil, QFilter). 
All gate lines were additionally filtered using 16~Hz low pass filters at room temperature.
The dc and ac currents were amplified using a commercially available current-to-voltage converter (Basel Precision Instruments) using a gain of $10^{6}$~V/A.
Both dc and ac voltages were amplified using commercial preamplifiers (Stanford Research, SR560) at a gain of $10^{3}$.
The microwave tone for Shapiro measurements was generated using a commercial microwave source (Rohde \& Schwarz SG100a), attenuated by 21~dB in the cryostat, and applied through a home-built $\lambda/4$ antenna directed at the sample.\\

\subsection{SHAPIRO STEPS}

The parametric representations of the Shapiro response depicted in Figs.~\ref{f4}(b) and \ref{f4}(d) were generated by plotting the differential resistance, $dV/dI$, of the rhombus as a function of the applied rf power, $P$, and the measured dc voltage, $V$, across the device in units of $h\nu/2e$.
Figure~\ref{fa2} shows a representative dc voltage map taken simultaneously with the $dV/dI$ map shown in Fig.~\ref{f4}(a).
The dc voltage shows well-developed Shapiro steps at multiples of $V = h\nu/2e$, corresponding to a charge-$2e$ supercurrent transport and $2\pi$-periodic CPR.

\subsection{RF RESPONSE OF AN UNBALANCED RHOMBUS}

The pattern of the observed Shapiro lobes contains information about the presence of different harmonics in the Josephson potential.
A current-phase relation containing a single harmonic produces a series of Shapiro steps, with step widths decreasing monotonically with increasing step number.
The harmonic determined the frequency of such Shapiro steps: a $\sin \varphi$ CPR results in integer Shapiro steps, while a $\sin 2\varphi$ CPR yields half-integer Shapiro steps.
When both $\sin \varphi$ and $\sin 2\varphi$ components are present with similar amplitudes, integer Shapiro steps become more pronounced than half-integer ones, resulting in an even-odd pattern in the Shapiro lobe size~\cite{Ueda2020Sep}.
This regime can be observed in the hybrid rhombus when $\sigma_1$ and $\sigma_2$ are tuned to be different, such that the $\sin \varphi$ component of the CPR is suppressed only partially; see Fig.~\ref{fa3}.

\bibliography{Literature}

\end{document}